\documentclass[amsmath,longbibliography,reprint,aip,apl]{revtex4-1}
\usepackage{hyperref}
\usepackage{graphicx}
\begin{document}

\title
{Waveguide coupled cavity-enhanced light emission from individual carbon nanotubes}

\author{D.~Yamashita}
\affiliation{Quantum Optoelectronics Research Team, RIKEN Center for Advanced Photonics, Saitama 351-0198, Japan}
\author{H.~Machiya}
\affiliation{Nanoscale Quantum Photonics Laboratory, RIKEN Cluster for Pioneering Research, Saitama 351-0198, Japan}
\affiliation{Department of Electrical Engineering, The University of Tokyo, Tokyo 113-8656, Japan}
\author{K.~Otsuka}
\affiliation{Nanoscale Quantum Photonics Laboratory, RIKEN Cluster for Pioneering Research, Saitama 351-0198, Japan}
\author{A.~Ishii}
\affiliation{Quantum Optoelectronics Research Team, RIKEN Center for Advanced Photonics, Saitama 351-0198, Japan}
\affiliation{Nanoscale Quantum Photonics Laboratory, RIKEN Cluster for Pioneering Research, Saitama 351-0198, Japan}
\author{Y.~K.~Kato}
\email[Corresponding author: ]{yuichiro.kato@riken.jp}
\affiliation{Quantum Optoelectronics Research Team, RIKEN Center for Advanced Photonics, Saitama 351-0198, Japan}
\affiliation{Nanoscale Quantum Photonics Laboratory, RIKEN Cluster for Pioneering Research, Saitama 351-0198, Japan}

\begin{abstract}
We demonstrate an individual single-walled carbon nanotube light emitter integrated onto a microcavity and a waveguide operating in the telecom wavelength regime. Light emission from the carbon nanotube is enhanced at the cavity resonance and is efficiently extracted from the waveguide facet. We have transferred carbon nanotubes to a nanobeam cavity with a dry process, ensuring that an individual carbon nanotube is used. The guided light emission from a chirality-identified single carbon nanotube has a narrow linewidth of less than 1.3~nm and an off-resonance rejection of $\sim$17~dB. The waveguide-coupled device configuration is compatible with fully integrated on-chip designs and is promising for carbon-nanotube-based photonics.
\end{abstract}

\maketitle

Silicon photonics has enabled on-chip integration of various optical components such as light sources, detectors, and optical modulators, expanding the capabilities of monolithic photonic circuits.~\cite{Thomson:2016, Wang:2017} For further scaling and increased functionalities, integration of nanoscale emitters is desirable. Since the low light emission efficiency of silicon is due to its indirect bandgap, \mbox{III--V} compound semiconductors with direct bandgaps are good alternatives as light sources.~\cite{Wang:2015, Chen:2016np} The compound semiconductors, however, cannot be grown directly on silicon substrates, which hinders their use in silicon photonics for a broad range of applications. 
In this regard, single-walled carbon nanotubes (CNTs) are compatible with silicon photonics because they can be grown directly on silicon substrates. Furthermore, the low-dimensionality effects on excitons give rise to unique electronic and optical properties.~\cite{Avouris:2008} CNTs exhibit telecom-band electroluminescence~\cite{Khasminskaya:2016,Higashide:2017} and photoluminescence (PL),~\cite{Lefebvre:2003} as well as single-photon emission at room temperature by utilizing localized exciton trapping sites~\cite{He:2018} and exciton-exciton annihilation process,~\cite{Khasminskaya:2016, Ishii:2017} which make them attractive light source candidates for silicon photonics. 

Although CNTs have numerous advantages, their quantum efficiencies are typically low~\cite{Miyauchi:2009, Crochet:2012} and they have broad spectral linewidths. To overcome these drawbacks, one of the most promising solutions is the use of silicon optical microcavities.~\cite{Watahiki:2012, Imamura:2013, Miura:2014, Noury:2015, Liu:2015, Pyatkov:2016, Hoang:2017, Biccari:2017, Machiya:2018, Ishii:2018, DuranValdeiglesias:2018, Zhang:2020, Higuchi:2020} The small mode volumes of the microcavities can increase Purcell enhancement and the high quality ($Q$) factors of the microcavities can also reduce the linewidth of the CNT emission. In addition, an optical waveguide can be easily integrated with these cavities,~\cite{Quan:2010,Akahane:2003Nature,Soltani:2007,Bogaerts:2012} connecting the light emitter and other optical components for mutual access. The integration of the waveguide and different optical components is thus crucial for on-chip photonic devices. Such waveguide-coupled and cavity-enhanced light emission from CNTs devices has been demonstrated using one- and two-dimensional photonic crystals (PCs)~\cite{Pyatkov:2016, Hoang:2017} as well as micro-ring and micro-disk~\cite{Noury:2015, DuranValdeiglesias:2018, Higuchi:2020} cavities.
In these studies, the deposition of CNTs, however, has been performed by dispersion~\cite{Noury:2015, Hoang:2017, DuranValdeiglesias:2018, Higuchi:2020} and dielectrophoresis~\cite{Pyatkov:2016} methods, which use solution processes. In order to harness the unique optical properties of CNTs such as single photon emission, it is important to isolate individual CNTs and eliminate contamination during the solution-based processes.

In this work, individual CNT telecom-wavelength emitters are integrated onto a microcavity and a waveguide. Using finite-difference time-domain (FDTD) simulations,~\cite{Oskooi:2010} we have designed an air-mode PC nanobeam cavity with one thin end mirror for guiding the light into the waveguide. CNTs are grown on a $\rm SiO_2$/Si substrate and transferred on the cavities through an all-dry process ensuring cleanliness of CNTs and devices. We characterize the devices using two geometries: a top detection configuration that measures light emission from CNTs on a nanobeam cavity, and a side detection configuration that collects light emission coupled to the waveguide. The nanobeam cavity with a small mode volume enhances light emission from a chirality-identified single CNT. The cavity-coupled light propagates into the waveguide and is emitted from the waveguide facet with a sharp linewidth and a large off-resonance rejection.

\begin{figure}[t]
\includegraphics[width=8.2cm]{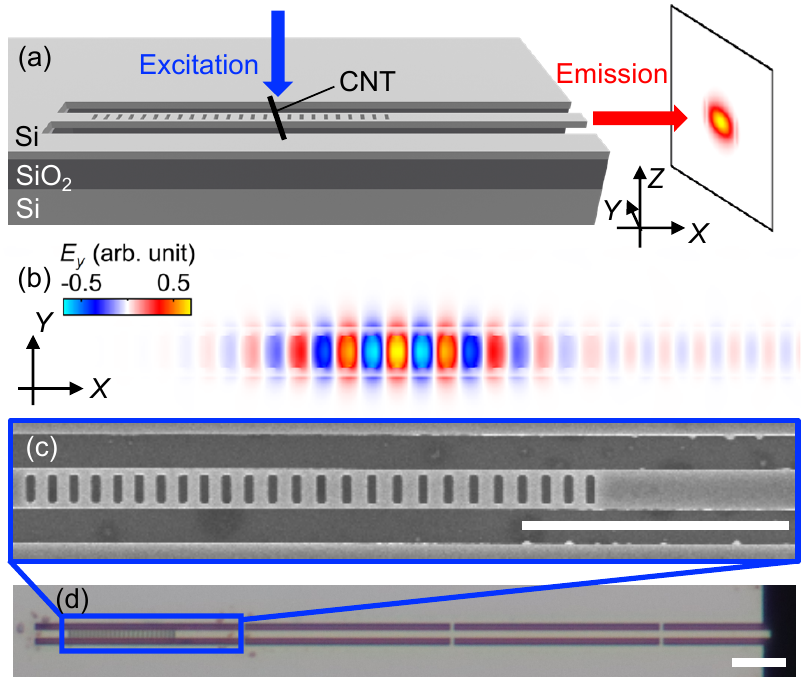}
\caption{
\label{Fig1} 
(a) Schematic of the device. The individual CNT located on the cavity is excited at normal incidence. The guided light is extracted from the waveguide facet. The 3D axis shows the sample direction.
(b) Simulated spatial distribution of the $y$-component of the electric field $E_{\rm y}$ for the fundamental transverse-electric mode at the surface of the substrate. The nanobeams are designed to be 800-nm-wide and the air-mode cavity has a lattice constant of $a$ = 350~nm. The cavities are designed to have increased lattice constant in a parabolic manner over 16 periods.~\cite{Miura:2014} The air-holes are 0.35$a$ $\times$ 500~nm and the cavity center has a period of 1.18$a$. The left side of the cavity center has 16 holes, which are sufficient to function as a Bragg mirror, while the right side has a fewer mirror number of 8 to facilitate light propagation. The trenches on both sides of the nanobeam are 600-nm-wide.
(c) Scanning electron micrograph and (d) optical image of a fabricated air-suspended nanobeam. The waveguides are supported by 600-nm-long beams every 20~$\mu$m on both sides. The scale bars in (c) and (d) are 5~$\mu$m. Panel (b) shares the same scale bar in (c). 
}
\end{figure}

Our CNT PL device schematic is illustrated in Fig.~\ref{Fig1}(a). We utilize air-mode nanobeam cavities having electric field maximum located in air for efficient coupling to air-suspended CNTs.~\cite{Miura:2014} The cavity with a small mode volume enables enhancement of light-matter interaction. To channel the coupled light into a waveguide, we adjust the number of air holes on one end of the cavity.~\cite{Pyatkov:2016} The cavity field profile is shown in Fig.~\ref{Fig1}(b). The field propagates into the waveguide, which would allow for extraction of narrow linewidth light emission from the waveguide facet. 

The air-mode nanobeam cavities and waveguides are fabricated on a silicon-on-insulator substrate with a 260-nm-thick top silicon layer and a 1-$\mu$m-thick buried-oxide (BOX) layer. After defining the PC pattern on a resist mask by electron beam lithography, the pattern is transferred to the top silicon slab through an inductively coupled plasma etching process using C$_4$F$_8$ and SF$_6$ gases. Following resist removal, the BOX layer is etched with 20\% hydrofluoric acid to form an air-suspended nanobeam structure. A scanning electron micrograph and an optical micrograph of a typical device are shown in Figs.~\ref{Fig1}(c) and \ref{Fig1}(d), respectively.

\begin{figure}[t]
\includegraphics[width=8.2cm]{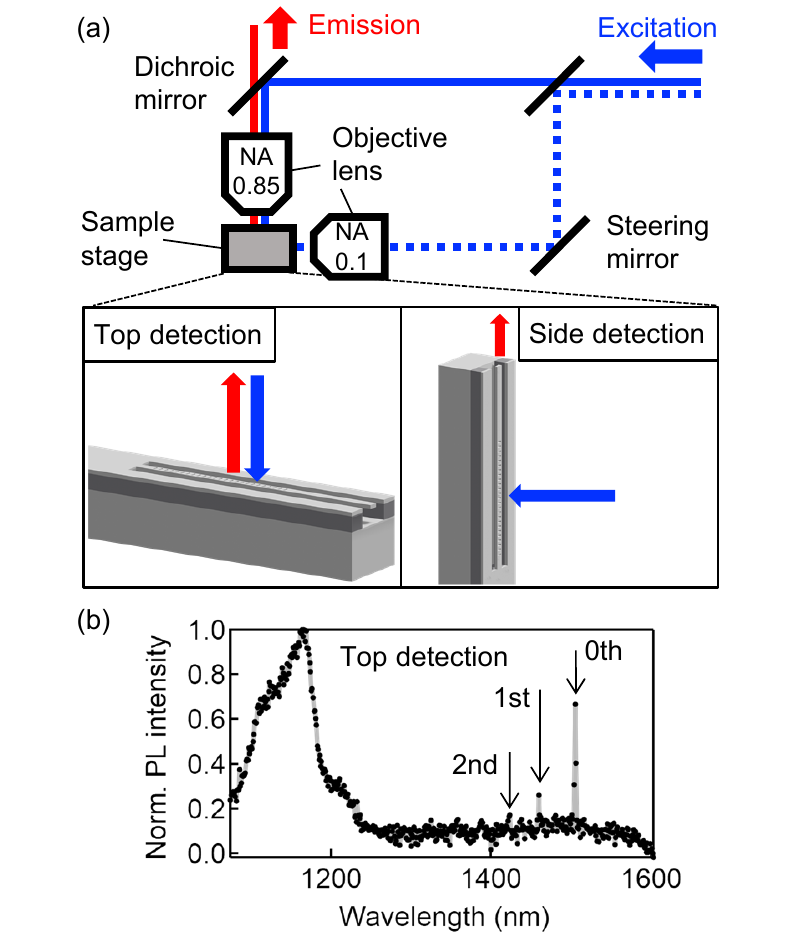}
\caption{
\label{Fig2} 
(a) Schematic of the setup. The excitation laser is irradiated on the surface of the sample through a path indicated by a blue solid (dashed) line for top (side) detection. The objective lens in the excitation path for the top (side) detection configuration has an NA of 0.85 (0.1) and a working distance of 1.2~mm (23~mm). The excitation laser spot size is 1~$\mu$m (8~$\mu$m) for the top (side) detection configuration.
(b) PL spectrum of silicon measured in the top detection configuration. The device shows a broad peak from silicon band edge emission as well as sharp peaks from the fundamental (0th) and high-order (1st and 2nd) cavity modes at resonant wavelengths of 1505.1, 1459.6 and 1423.4~nm, for the 0th, 1st, and 2nd modes. An excitation power of 200~$\mu$W and an excitation wavelength of 780~nm are used. The laser polarization is perpendicular to the nanobeam. All measurements are performed at room temperature in ambient condition.
}
\end{figure}

\begin{figure}[t]
\includegraphics[width=8.2cm]{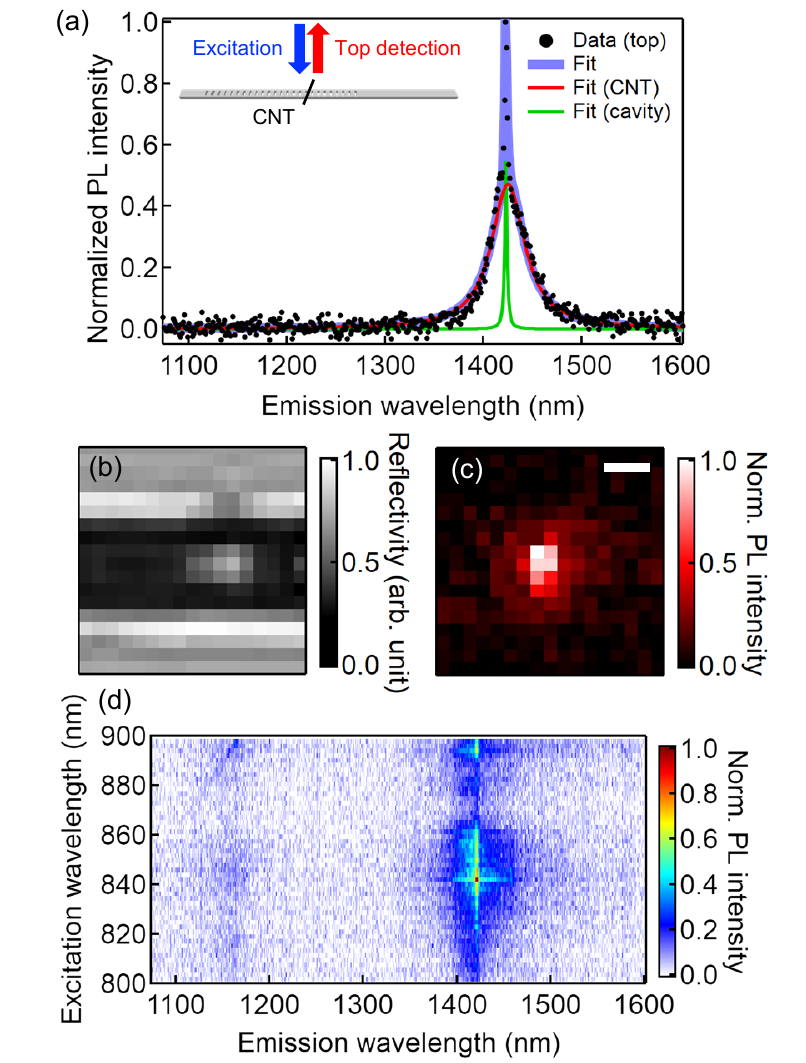}
\caption{
\label{Fig3}
(a) PL spectrum of a CNT coupled to a cavity measured in the top detection configuration as illustrated in the inset. The spectrum is taken at a position where the cavity emission is maximized. The black dots are data and the blue line is a bi-Lorentzian fit. The red and green curves correspond to the CNT and the cavity peak components, respectively. The CNT peak is centered at 1424.9~nm and has a linewidth of 41.1$\pm$1.0~nm, whereas the cavity peak has a center wavelength of 1422.7~nm with a linewidth of 2.6~nm.
(b) Reflectivity image around the nanobeam and the CNT. 
(c) PL image in the same area extracted at the wavelength of the cavity emission peak over a 20-nm-wide spectral window. The scale bar in (c) is 1~$\mu$m and is shared with panel (b). 
(d) PL excitation map. For (a)--(c), the excitation laser wavelength is 844~nm and the laser polarization is parallel to the CNT. (a)--(d) are taken with an excitation power of 100~$\mu$W.
}
\end{figure}

\begin{figure}[t]
\includegraphics[width=8.2cm]{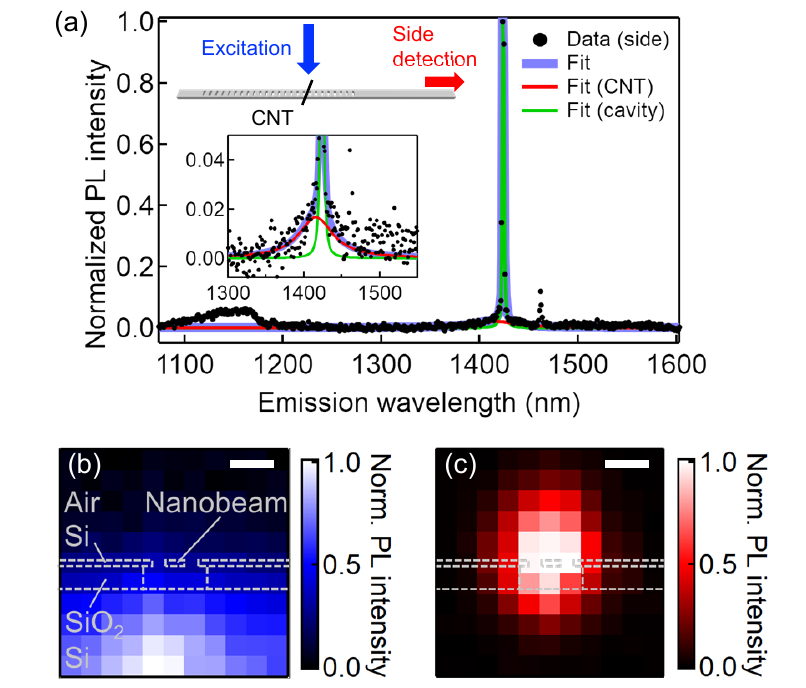}
\caption{
\label{Fig4}
(a) PL spectrum of the CNT coupled to a cavity measured in the side detection configuration as illustrated in the upper inset. The spectrum is taken at a position where the cavity emission is maximized. The black dots are data and the blue line is a bi-Lorentzian fit. The red and green curves correspond to the CNT and the cavity peak components, respectively. The CNT peak is centered at 1415.8~nm and has a linewidth of 54.5$\pm$15.8~nm, whereas the cavity peak has a center wavelength of 1423.9~nm with a linewidth of 1.3~nm. The plot inset is an enlarged view of the low-intensity region around the cavity emission peak. 
(b) and (c) Waveguided PL images extracted at the wavelengths of the silicon emission peak at 1152.4~nm and the cavity emission peak, respectively. PL intensity is obtained by integrating PL over a 1-nm-wide spectral window. The scale bars are 2~$\mu$m. The outline of the device cross-section is overlaid. (b) and (c) are obtained by scanning the 3D stage. For all PL measurements, an excitation power of 2~mW and an excitation wavelength of 844~nm are used  whereas the laser polarization is parallel to the CNT.
}
\end{figure}

Before transferring CNTs, we characterize the resonant modes of the fabricated nanobeam cavities. The devices are measured using a confocal microscope system as shown in Fig.~\ref{Fig2}(a). We use a wavelength-tunable Ti:sapphire laser for excitation with its power and polarization angle controlled by neutral density filters and a half-wave plate, respectively. The excitation laser is irradiated on the surface of the sample through a path indicated by the blue solid line for top detection. We mount the sample on a motorized three-dimensional (3D) feedback stage in the top detection configuration shown in the bottom left panel in Fig.~\ref{Fig2}(a). PL is collected using the objective lens with a numerical aperture (NA) of 0.85 and is detected using a liquid-nitrogen-cooled \mbox{InGaAs} diode array attached to a spectrometer. The laser reflection comes back through the objective lens and is detected with a silicon photodiode, which is used to align to the cavity position. 

Figure~\ref{Fig2}(b) shows a silicon PL spectrum measured before transferring CNTs. The device shows sharp resonant mode peaks in the telecom wavelength regime. These peaks are near the emission wavelengths of the CNT chiralities with large abundance from our synthesis,~\cite{Ishii:2015} which maximizes the coupling probability between CNT emission and the cavity modes. Although the $Q$ factor decreases for the higher order modes, the linewidth of cavity modes is still sufficiently narrow compared to a typical emission linewidth of a single CNT. 

To couple individual CNTs to a nanobeam cavity, we utilize an anthracene-assisted dry transfer method.~\cite{Otsuka:2020} The use of anthracene crystals allows transfer of CNTs onto arbitrary substrates, and sublimation of anthracene leaves behind clean CNTs onto the nanobeam cavities because contamination from solvents is absent in the all-dry process. We transfer CNTs grown on a $\rm SiO_2$/Si substrate onto the cavities, and individual CNTs that are optically coupled are identified by PL scans over the nanobeams. Since the cavity linewidth is considerably narrower than the CNT linewidth at room temperature, devices showing a sharp peak component indicate that CNT emission is coupled to the cavity and are chosen for detailed characterization.

Figure~\ref{Fig3}(a) shows a PL spectrum of a CNT coupled to a cavity. On top of the broad direct emission from the CNT, there is a sharp peak corresponding to the cavity mode. We use a bi-Lorentzian function to decompose the spectrum into the cavity mode (green) and the broad nanotube emission (red). The $Q$ factor of the cavity is estimated to be 540 from the spectral linewidth of 2.6~nm. A comparison with the resonant mode peaks [Fig.~\ref{Fig2}(b)] shows that the CNT emission couples to the 2nd mode of the cavity. The coupling factor~$\beta$ is estimated to be 0.48 from the spectral intensity ratio between the CNT emission and cavity mode, taking into account the different collection efficiencies for the CNT emission (42\%) and the 2nd cavity mode (3.5\%) as calculated from FDTD simulations. Figures~\ref{Fig3}(b) and \ref{Fig3}(c) shows a reflectivity image and a PL image extracted at the wavelength of the cavity emission peak, respectively. We observe only a single spot that is within the cavity mode in Fig.~\ref{Fig3}(c), indicating the CNT is spatially localized. To confirm that the CNT is individual, a PL excitation map is also taken [Fig.~\ref{Fig3}(d)]. A single main peak in the map confirms that the cavity is coupled to an individual CNT, and its chirality is assigned to be (10,8) by comparing to tabulated data.~\cite{Ishii:2015}

For the device coupled with a single CNT, we investigate the PL emitted from the nanobeam waveguide facet. The sample is mounted on the stage in the side detection configuration shown in the bottom right panel of Fig.~\ref{Fig2}(a). The excitation laser is irradiated on the surface of the sample through a path indicated by the blue dashed line in Fig.~\ref{Fig2}(a) and the excitation spot is scanned by a steering mirror. Figure~\ref{Fig4}(a) shows a waveguided CNT PL spectrum of the same device shown in Fig.~\ref{Fig3}. The light emission from the identified CNT is enhanced at the cavity resonance and extracted from the waveguide facet. Compared to the spectrum in Fig.~\ref{Fig3}(a), the broad CNT component is suppressed due to the PC bandgap. The intensity ratio of the uncoupled CNT peak to the cavity mode peak is 0.022, which corresponds to an off-resonance rejection of $\sim$17~dB. The coupled PL emission has a narrow linewidth of less than 1.3~nm, which is slightly smaller than that measured by using the top detection configuration [Fig.~\ref{Fig3}(a)]. The $Q$ factor of the cavity is estimated to be 1100 from the spectral linewidth and is limited by the spectrometer resolution of $\sim$1~nm.

Figures~\ref{Fig4}(b) and \ref{Fig4}(c) are waveguided PL images extracted at the wavelengths of a silicon emission peak and a cavity emission peak, respectively. The silicon PL is strong below the substrate surface [Fig.~\ref{Fig4}(b)], whereas the cavity PL is centered at the waveguide facet [Fig.~\ref{Fig4}(c)]. This spatial localization of PL emission clearly indicates that the CNT emission coupled to the cavity propagates through the nanobeam waveguide. 
Collection efficiency of the light emitted from the waveguide facet is much higher than that from the top of the cavity, since most of the light coupled to the cavity goes through the waveguide. It should be possible to achieve further PL enhancement and effective PL extraction by deterministically placing CNTs\cite{Otsuka:2020} in the center of the cavities, optimizing the cavity structures,~\cite{Quan:2011} and introducing a tapered structure at the waveguide facet.~\cite{Shani:1989}

In conclusion, we have demonstrated narrow-linewidth and low-background light emission from a single CNT integrated onto a microcavity and a waveguide operating in the telecom E-band. The waveguided light spectrum has a sharp cavity-enhanced peak with a less than 1.3-nm linewidth and a large off-resonance rejection of $\sim$17~dB. 
The use of a dry transfer method allows integration of clean CNTs and photonic devices, where the unique optical properties of CNTs can be utilized for applications such as single photon emission. The waveguide-coupled light can easily be connected to various optical components on a monolithic chip and optical fibers~\cite{Tiecke:2015, JimenezGordillo:2019}. Our device configuration opens a new path toward CNT-based photonic devices integrated with microcavities and waveguides. 

\begin{acknowledgments}
This work is supported in part by JSPS (KAKENHI JP19J00894, JP20H02558, JP20J00817, JP20K15137, JP20K15199), MIC (SCOPE 191503001), MEXT (Nanotechnology Platform JPMXP09F19UT0078) and RIKEN (Incentive Research Project). D.Y., H.M. and K.O. are supported by the JSPS Research Fellowship. The FDTD calculations were performed using the HOKUSAI BigWaterfall supercomputer at RIKEN. We acknowledge the Advanced Manufacturing Support Team at RIKEN for technical assistance.
\end{acknowledgments}

\section*{Data Availability}
The data that support the findings of this study are available from the corresponding author upon reasonable request.


\begin{thebibliography}{36}%
\makeatletter
\providecommand \@ifxundefined [1]{%
 \@ifx{#1\undefined}
}%
\providecommand \@ifnum [1]{%
 \ifnum #1\expandafter \@firstoftwo
 \else \expandafter \@secondoftwo
 \fi
}%
\providecommand \@ifx [1]{%
 \ifx #1\expandafter \@firstoftwo
 \else \expandafter \@secondoftwo
 \fi
}%
\providecommand \natexlab [1]{#1}%
\providecommand \enquote  [1]{``#1''}%
\providecommand \bibnamefont  [1]{#1}%
\providecommand \bibfnamefont [1]{#1}%
\providecommand \citenamefont [1]{#1}%
\providecommand \href@noop [0]{\@secondoftwo}%
\providecommand \href [0]{\begingroup \@sanitize@url \@href}%
\providecommand \@href[1]{\@@startlink{#1}\@@href}%
\providecommand \@@href[1]{\endgroup#1\@@endlink}%
\providecommand \@sanitize@url [0]{\catcode `\\12\catcode `\$12\catcode
  `\&12\catcode `\#12\catcode `\^12\catcode `\_12\catcode `\%12\relax}%
\providecommand \@@startlink[1]{}%
\providecommand \@@endlink[0]{}%
\providecommand \url  [0]{\begingroup\@sanitize@url \@url }%
\providecommand \@url [1]{\endgroup\@href {#1}{\urlprefix }}%
\providecommand \urlprefix  [0]{URL }%
\providecommand \Eprint [0]{\href }%
\providecommand \doibase [0]{http://dx.doi.org/}%
\providecommand \selectlanguage [0]{\@gobble}%
\providecommand \bibinfo  [0]{\@secondoftwo}%
\providecommand \bibfield  [0]{\@secondoftwo}%
\providecommand \translation [1]{[#1]}%
\providecommand \BibitemOpen [0]{}%
\providecommand \bibitemStop [0]{}%
\providecommand \bibitemNoStop [0]{.\EOS\space}%
\providecommand \EOS [0]{\spacefactor3000\relax}%
\providecommand \BibitemShut  [1]{\csname bibitem#1\endcsname}%
\let\auto@bib@innerbib\@empty
\bibitem [{\citenamefont {Thomson}\ \emph {et~al.}(2016)\citenamefont
  {Thomson}, \citenamefont {Zilkie}, \citenamefont {Bowers}, \citenamefont
  {Komljenovic}, \citenamefont {Reed}, \citenamefont {Vivien}, \citenamefont
  {Marris-Morini}, \citenamefont {Cassan}, \citenamefont {Virot}, \citenamefont
  {F{\'e}d{\'e}li}, \citenamefont {Hartmann}, \citenamefont {Schmid},
  \citenamefont {Xu}, \citenamefont {Boeuf}, \citenamefont {O'Brien},
  \citenamefont {Mashanovich},\ and\ \citenamefont
  {Nedeljkovic}}]{Thomson:2016}%
  \BibitemOpen
  \bibfield  {author} {\bibinfo {author} {\bibfnamefont {D.}~\bibnamefont
  {Thomson}}, \bibinfo {author} {\bibfnamefont {A.}~\bibnamefont {Zilkie}},
  \bibinfo {author} {\bibfnamefont {J.~E.}\ \bibnamefont {Bowers}}, \bibinfo
  {author} {\bibfnamefont {T.}~\bibnamefont {Komljenovic}}, \bibinfo {author}
  {\bibfnamefont {G.~T.}\ \bibnamefont {Reed}}, \bibinfo {author}
  {\bibfnamefont {L.}~\bibnamefont {Vivien}}, \bibinfo {author} {\bibfnamefont
  {D.}~\bibnamefont {Marris-Morini}}, \bibinfo {author} {\bibfnamefont
  {E.}~\bibnamefont {Cassan}}, \bibinfo {author} {\bibfnamefont
  {L.}~\bibnamefont {Virot}}, \bibinfo {author} {\bibfnamefont {J.-M.}\
  \bibnamefont {F{\'e}d{\'e}li}}, \bibinfo {author} {\bibfnamefont {J.-M.}\
  \bibnamefont {Hartmann}}, \bibinfo {author} {\bibfnamefont {J.~H.}\
  \bibnamefont {Schmid}}, \bibinfo {author} {\bibfnamefont {D.-X.}\
  \bibnamefont {Xu}}, \bibinfo {author} {\bibfnamefont {F.}~\bibnamefont
  {Boeuf}}, \bibinfo {author} {\bibfnamefont {P.}~\bibnamefont {O'Brien}},
  \bibinfo {author} {\bibfnamefont {G.~Z.}\ \bibnamefont {Mashanovich}}, \ and\
  \bibinfo {author} {\bibfnamefont {M.}~\bibnamefont {Nedeljkovic}},\
  }\bibfield  {title} {\bibinfo {title} {Roadmap on silicon photonics},\ }\href
  {\doibase 10.1088/2040-8978/18/7/073003} {\bibfield  {journal} {\bibinfo
  {journal} {J. Opt.}\ }\textbf {\bibinfo {volume} {18}},\ \bibinfo {eid}
  {073003} (\bibinfo {year} {2016})}\BibitemShut {NoStop}%
\bibitem [{\citenamefont {Wang}\ \emph {et~al.}(2017)\citenamefont {Wang},
  \citenamefont {Abbasi}, \citenamefont {Dave}, \citenamefont {De~Groote},
  \citenamefont {Kumari}, \citenamefont {Kunert}, \citenamefont {Merckling},
  \citenamefont {Pantouvaki}, \citenamefont {Shi}, \citenamefont {Tian},
  \citenamefont {Van~Gasse}, \citenamefont {Verbist}, \citenamefont {Wang},
  \citenamefont {Xie}, \citenamefont {Zhang}, \citenamefont {Zhu},
  \citenamefont {Bauwelinck}, \citenamefont {Yin}, \citenamefont {Hens},
  \citenamefont {Van~Campenhout}, \citenamefont {Kuyken}, \citenamefont
  {Baets}, \citenamefont {Morthier}, \citenamefont {Van~Thourhout},\ and\
  \citenamefont {Roelkens}}]{Wang:2017}%
  \BibitemOpen
  \bibfield  {author} {\bibinfo {author} {\bibfnamefont {Z.}~\bibnamefont
  {Wang}}, \bibinfo {author} {\bibfnamefont {A.}~\bibnamefont {Abbasi}},
  \bibinfo {author} {\bibfnamefont {U.}~\bibnamefont {Dave}}, \bibinfo {author}
  {\bibfnamefont {A.}~\bibnamefont {De~Groote}}, \bibinfo {author}
  {\bibfnamefont {S.}~\bibnamefont {Kumari}}, \bibinfo {author} {\bibfnamefont
  {B.}~\bibnamefont {Kunert}}, \bibinfo {author} {\bibfnamefont
  {C.}~\bibnamefont {Merckling}}, \bibinfo {author} {\bibfnamefont
  {M.}~\bibnamefont {Pantouvaki}}, \bibinfo {author} {\bibfnamefont
  {Y.}~\bibnamefont {Shi}}, \bibinfo {author} {\bibfnamefont {B.}~\bibnamefont
  {Tian}}, \bibinfo {author} {\bibfnamefont {K.}~\bibnamefont {Van~Gasse}},
  \bibinfo {author} {\bibfnamefont {J.}~\bibnamefont {Verbist}}, \bibinfo
  {author} {\bibfnamefont {R.}~\bibnamefont {Wang}}, \bibinfo {author}
  {\bibfnamefont {W.}~\bibnamefont {Xie}}, \bibinfo {author} {\bibfnamefont
  {J.}~\bibnamefont {Zhang}}, \bibinfo {author} {\bibfnamefont
  {Y.}~\bibnamefont {Zhu}}, \bibinfo {author} {\bibfnamefont {J.}~\bibnamefont
  {Bauwelinck}}, \bibinfo {author} {\bibfnamefont {X.}~\bibnamefont {Yin}},
  \bibinfo {author} {\bibfnamefont {Z.}~\bibnamefont {Hens}}, \bibinfo {author}
  {\bibfnamefont {J.}~\bibnamefont {Van~Campenhout}}, \bibinfo {author}
  {\bibfnamefont {B.}~\bibnamefont {Kuyken}}, \bibinfo {author} {\bibfnamefont
  {R.}~\bibnamefont {Baets}}, \bibinfo {author} {\bibfnamefont
  {G.}~\bibnamefont {Morthier}}, \bibinfo {author} {\bibfnamefont
  {D.}~\bibnamefont {Van~Thourhout}}, \ and\ \bibinfo {author} {\bibfnamefont
  {G.}~\bibnamefont {Roelkens}},\ }\bibfield  {title} {\bibinfo {title} {Novel
  light source integration approaches for silicon photonics},\ }\href {\doibase
  10.1002/lpor.201700063} {\bibfield  {journal} {\bibinfo  {journal} {Laser
  Photon. Rev.}\ }\textbf {\bibinfo {volume} {11}},\ \bibinfo {pages} {1700063}
  (\bibinfo {year} {2017})}\BibitemShut {NoStop}%
\bibitem [{\citenamefont {Wang}\ \emph {et~al.}(2015)\citenamefont {Wang},
  \citenamefont {Tian}, \citenamefont {Pantouvaki}, \citenamefont {Guo},
  \citenamefont {Absil}, \citenamefont {van Campenhout}, \citenamefont
  {Merckling},\ and\ \citenamefont {van Thourhout}}]{Wang:2015}%
  \BibitemOpen
  \bibfield  {author} {\bibinfo {author} {\bibfnamefont {Z.}~\bibnamefont
  {Wang}}, \bibinfo {author} {\bibfnamefont {B.}~\bibnamefont {Tian}}, \bibinfo
  {author} {\bibfnamefont {M.}~\bibnamefont {Pantouvaki}}, \bibinfo {author}
  {\bibfnamefont {W.}~\bibnamefont {Guo}}, \bibinfo {author} {\bibfnamefont
  {P.}~\bibnamefont {Absil}}, \bibinfo {author} {\bibfnamefont
  {J.}~\bibnamefont {van Campenhout}}, \bibinfo {author} {\bibfnamefont
  {C.}~\bibnamefont {Merckling}}, \ and\ \bibinfo {author} {\bibfnamefont
  {D.}~\bibnamefont {van Thourhout}},\ }\bibfield  {title} {\bibinfo {title}
  {Room-temperature \mbox{InP} distributed feedback laser array directly grown on
  silicon},\ }\href {\doibase 10.1038/nphoton.2015.199} {\bibfield  {journal}
  {\bibinfo  {journal} {Nat. Photon.}\ }\textbf {\bibinfo {volume} {9}},\
  \bibinfo {pages} {837} (\bibinfo {year} {2015})}\BibitemShut {NoStop}%
\bibitem [{\citenamefont {Chen}\ \emph {et~al.}(2016)\citenamefont {Chen},
  \citenamefont {Li}, \citenamefont {Wu}, \citenamefont {Jiang}, \citenamefont
  {Tang}, \citenamefont {Shutts}, \citenamefont {Elliott}, \citenamefont
  {Sobiesierski}, \citenamefont {Seeds}, \citenamefont {Ross}, \citenamefont
  {Smowton},\ and\ \citenamefont {Liu}}]{Chen:2016np}%
  \BibitemOpen
  \bibfield  {author} {\bibinfo {author} {\bibfnamefont {S.}~\bibnamefont
  {Chen}}, \bibinfo {author} {\bibfnamefont {W.}~\bibnamefont {Li}}, \bibinfo
  {author} {\bibfnamefont {J.}~\bibnamefont {Wu}}, \bibinfo {author}
  {\bibfnamefont {Q.}~\bibnamefont {Jiang}}, \bibinfo {author} {\bibfnamefont
  {M.}~\bibnamefont {Tang}}, \bibinfo {author} {\bibfnamefont {S.}~\bibnamefont
  {Shutts}}, \bibinfo {author} {\bibfnamefont {S.~N.}\ \bibnamefont {Elliott}},
  \bibinfo {author} {\bibfnamefont {A.}~\bibnamefont {Sobiesierski}}, \bibinfo
  {author} {\bibfnamefont {A.~J.}\ \bibnamefont {Seeds}}, \bibinfo {author}
  {\bibfnamefont {I.}~\bibnamefont {Ross}}, \bibinfo {author} {\bibfnamefont
  {P.~M.}\ \bibnamefont {Smowton}}, \ and\ \bibinfo {author} {\bibfnamefont
  {H.}~\bibnamefont {Liu}},\ }\bibfield  {title} {\bibinfo {title}
  {Electrically pumped continuous-wave \mbox{III--V} quantum dot lasers on silicon},\
  }\href {\doibase 10.1038/nphoton.2016.21} {\bibfield  {journal} {\bibinfo
  {journal} {Nat. Photon.}\ }\textbf {\bibinfo {volume} {10}},\ \bibinfo
  {pages} {307} (\bibinfo {year} {2016})}\BibitemShut {NoStop}%
\bibitem [{\citenamefont {Avouris}\ \emph {et~al.}(2008)\citenamefont
  {Avouris}, \citenamefont {Freitag},\ and\ \citenamefont
  {Perebeinos}}]{Avouris:2008}%
  \BibitemOpen
  \bibfield  {author} {\bibinfo {author} {\bibfnamefont {P.}~\bibnamefont
  {Avouris}}, \bibinfo {author} {\bibfnamefont {M.}~\bibnamefont {Freitag}}, \
  and\ \bibinfo {author} {\bibfnamefont {V.}~\bibnamefont {Perebeinos}},\
  }\bibfield  {title} {\bibinfo {title} {Carbon-nanotube photonics and
  optoelectronics},\ }\href {\doibase 10.1038/nphoton.2008.94} {\bibfield
  {journal} {\bibinfo  {journal} {Nat. Photon.}\ }\textbf {\bibinfo {volume}
  {2}},\ \bibinfo {pages} {341} (\bibinfo {year} {2008})}\BibitemShut {NoStop}%
\bibitem [{\citenamefont {Khasminskaya}\ \emph {et~al.}(2016)\citenamefont
  {Khasminskaya}, \citenamefont {Pyatkov}, \citenamefont {S\l{}owik},
  \citenamefont {Ferrari}, \citenamefont {Kahl}, \citenamefont {Kovalyuk},
  \citenamefont {Rath}, \citenamefont {Vetter}, \citenamefont {Hennrich},
  \citenamefont {Kappes}, \citenamefont {Gol'tsman}, \citenamefont {Korneev},
  \citenamefont {Rockstuhl}, \citenamefont {Krupke},\ and\ \citenamefont
  {Pernice}}]{Khasminskaya:2016}%
  \BibitemOpen
  \bibfield  {author} {\bibinfo {author} {\bibfnamefont {S.}~\bibnamefont
  {Khasminskaya}}, \bibinfo {author} {\bibfnamefont {F.}~\bibnamefont
  {Pyatkov}}, \bibinfo {author} {\bibfnamefont {K.}~\bibnamefont {S\l{}owik}},
  \bibinfo {author} {\bibfnamefont {S.}~\bibnamefont {Ferrari}}, \bibinfo
  {author} {\bibfnamefont {O.}~\bibnamefont {Kahl}}, \bibinfo {author}
  {\bibfnamefont {V.}~\bibnamefont {Kovalyuk}}, \bibinfo {author}
  {\bibfnamefont {P.}~\bibnamefont {Rath}}, \bibinfo {author} {\bibfnamefont
  {A.}~\bibnamefont {Vetter}}, \bibinfo {author} {\bibfnamefont
  {F.}~\bibnamefont {Hennrich}}, \bibinfo {author} {\bibfnamefont {M.~M.}\
  \bibnamefont {Kappes}}, \bibinfo {author} {\bibfnamefont {G.}~\bibnamefont
  {Gol'tsman}}, \bibinfo {author} {\bibfnamefont {A.}~\bibnamefont {Korneev}},
  \bibinfo {author} {\bibfnamefont {C.}~\bibnamefont {Rockstuhl}}, \bibinfo
  {author} {\bibfnamefont {R.}~\bibnamefont {Krupke}}, \ and\ \bibinfo {author}
  {\bibfnamefont {W.~H.~P.}\ \bibnamefont {Pernice}},\ }\bibfield  {title}
  {\bibinfo {title} {Fully integrated quantum photonic circuit with an
  electrically driven light source},\ }\href {\doibase
  10.1038/nphoton.2016.178} {\bibfield  {journal} {\bibinfo  {journal} {Nat.
  Photon.}\ }\textbf {\bibinfo {volume} {10}},\ \bibinfo {pages} {727}
  (\bibinfo {year} {2016})}\BibitemShut {NoStop}%
\bibitem [{\citenamefont {Higashide}\ \emph {et~al.}(2017)\citenamefont
  {Higashide}, \citenamefont {Yoshida}, \citenamefont {Uda}, \citenamefont
  {Ishii},\ and\ \citenamefont {Kato}}]{Higashide:2017}%
  \BibitemOpen
  \bibfield  {author} {\bibinfo {author} {\bibfnamefont {N.}~\bibnamefont
  {Higashide}}, \bibinfo {author} {\bibfnamefont {M.}~\bibnamefont {Yoshida}},
  \bibinfo {author} {\bibfnamefont {T.}~\bibnamefont {Uda}}, \bibinfo {author}
  {\bibfnamefont {A.}~\bibnamefont {Ishii}}, \ and\ \bibinfo {author}
  {\bibfnamefont {Y.~K.}\ \bibnamefont {Kato}},\ }\bibfield  {title} {\bibinfo
  {title} {Cold exciton electroluminescence from air-suspended carbon nanotube
  split-gate devices},\ }\href {\doibase 10.1063/1.4983278} {\bibfield
  {journal} {\bibinfo  {journal} {Appl. Phys. Lett.}\ }\textbf {\bibinfo
  {volume} {110}},\ \bibinfo {pages} {191101} (\bibinfo {year}
  {2017})}\BibitemShut {NoStop}%
\bibitem [{\citenamefont {Lefebvre}\ \emph {et~al.}(2003)\citenamefont
  {Lefebvre}, \citenamefont {Homma},\ and\ \citenamefont
  {Finnie}}]{Lefebvre:2003}%
  \BibitemOpen
  \bibfield  {author} {\bibinfo {author} {\bibfnamefont {J.}~\bibnamefont
  {Lefebvre}}, \bibinfo {author} {\bibfnamefont {Y.}~\bibnamefont {Homma}}, \
  and\ \bibinfo {author} {\bibfnamefont {P.}~\bibnamefont {Finnie}},\
  }\bibfield  {title} {\bibinfo {title} {Bright band gap photoluminescence from
  unprocessed single-walled carbon nanotubes},\ }\href {\doibase
  10.1103/PhysRevLett.90.217401} {\bibfield  {journal} {\bibinfo  {journal}
  {Phys. Rev. Lett.}\ }\textbf {\bibinfo {volume} {90}},\ \bibinfo {pages}
  {217401} (\bibinfo {year} {2003})}\BibitemShut {NoStop}%
\bibitem [{\citenamefont {He}\ \emph {et~al.}(2018)\citenamefont {He},
  \citenamefont {Htoon}, \citenamefont {Doorn}, \citenamefont {Pernice},
  \citenamefont {Pyatkov}, \citenamefont {Krupke}, \citenamefont {Jeantet},
  \citenamefont {Chassagneux},\ and\ \citenamefont {Voisin}}]{He:2018}%
  \BibitemOpen
  \bibfield  {author} {\bibinfo {author} {\bibfnamefont {X.}~\bibnamefont
  {He}}, \bibinfo {author} {\bibfnamefont {H.}~\bibnamefont {Htoon}}, \bibinfo
  {author} {\bibfnamefont {S.~K.}\ \bibnamefont {Doorn}}, \bibinfo {author}
  {\bibfnamefont {W.~H.~P.}\ \bibnamefont {Pernice}}, \bibinfo {author}
  {\bibfnamefont {F.}~\bibnamefont {Pyatkov}}, \bibinfo {author} {\bibfnamefont
  {R.}~\bibnamefont {Krupke}}, \bibinfo {author} {\bibfnamefont
  {A.}~\bibnamefont {Jeantet}}, \bibinfo {author} {\bibfnamefont
  {Y.}~\bibnamefont {Chassagneux}}, \ and\ \bibinfo {author} {\bibfnamefont
  {C.}~\bibnamefont {Voisin}},\ }\bibfield  {title} {\bibinfo {title} {Carbon
  nanotubes as emerging quantum-light sources},\ }\href {\doibase
  10.1038/s41563-018-0109-2} {\bibfield  {journal} {\bibinfo  {journal} {Nat.
  Mater.}\ }\textbf {\bibinfo {volume} {17}},\ \bibinfo {pages} {663}
  (\bibinfo {year} {2018})}\BibitemShut {NoStop}%
\bibitem [{\citenamefont {Ishii}\ \emph {et~al.}(2017)\citenamefont {Ishii},
  \citenamefont {Uda},\ and\ \citenamefont {Kato}}]{Ishii:2017}%
  \BibitemOpen
  \bibfield  {author} {\bibinfo {author} {\bibfnamefont {A.}~\bibnamefont
  {Ishii}}, \bibinfo {author} {\bibfnamefont {T.}~\bibnamefont {Uda}}, \ and\
  \bibinfo {author} {\bibfnamefont {Y.~K.}\ \bibnamefont {Kato}},\ }\bibfield
  {title} {\bibinfo {title} {Room-temperature single-photon emission from
  micrometer-long air-suspended carbon nanotubes},\ }\href {\doibase
  10.1103/physrevapplied.8.054039} {\bibfield  {journal} {\bibinfo  {journal}
  {Phys. Rev. Applied}\ }\textbf {\bibinfo {volume} {8}},\ \bibinfo {pages}
  {054039} (\bibinfo {year} {2017})}\BibitemShut {NoStop}%
\bibitem [{\citenamefont {Miyauchi}\ \emph {et~al.}(2009)\citenamefont
  {Miyauchi}, \citenamefont {Hirori}, \citenamefont {Matsuda},\ and\
  \citenamefont {Kanemitsu}}]{Miyauchi:2009}%
  \BibitemOpen
  \bibfield  {author} {\bibinfo {author} {\bibfnamefont {Y.}~\bibnamefont
  {Miyauchi}}, \bibinfo {author} {\bibfnamefont {H.}~\bibnamefont {Hirori}},
  \bibinfo {author} {\bibfnamefont {K.}~\bibnamefont {Matsuda}}, \ and\
  \bibinfo {author} {\bibfnamefont {Y.}~\bibnamefont {Kanemitsu}},\ }\bibfield
  {title} {\bibinfo {title} {Radiative lifetimes and coherence lengths of
  one-dimensional excitons in single-walled carbon nanotubes},\ }\href
  {\doibase 10.1103/PhysRevB.80.081410} {\bibfield  {journal} {\bibinfo
  {journal} {Phys. Rev. B}\ }\textbf {\bibinfo {volume} {80}},\ \bibinfo
  {pages} {081410(R)} (\bibinfo {year} {2009})}\BibitemShut {NoStop}%
\bibitem [{\citenamefont {Crochet}\ \emph {et~al.}(2012)\citenamefont
  {Crochet}, \citenamefont {Duque}, \citenamefont {Werner}, \citenamefont
  {Lounis}, \citenamefont {Cognet},\ and\ \citenamefont
  {Doorn}}]{Crochet:2012}%
  \BibitemOpen
  \bibfield  {author} {\bibinfo {author} {\bibfnamefont {J.~J.}\ \bibnamefont
  {Crochet}}, \bibinfo {author} {\bibfnamefont {J.~G.}\ \bibnamefont {Duque}},
  \bibinfo {author} {\bibfnamefont {J.~H.}\ \bibnamefont {Werner}}, \bibinfo
  {author} {\bibfnamefont {B.}~\bibnamefont {Lounis}}, \bibinfo {author}
  {\bibfnamefont {L.}~\bibnamefont {Cognet}}, \ and\ \bibinfo {author}
  {\bibfnamefont {S.~K.}\ \bibnamefont {Doorn}},\ }\bibfield  {title} {\bibinfo
  {title} {Disorder limited exciton transport in colloidal single-wall carbon
  nanotubes},\ }\href {\doibase 10.1021/nl301739d} {\bibfield  {journal}
  {\bibinfo  {journal} {Nano Lett.}\ }\textbf {\bibinfo {volume} {12}},\
  \bibinfo {pages} {5091} (\bibinfo {year} {2012})}\BibitemShut {NoStop}%
\bibitem [{\citenamefont {Watahiki}\ \emph {et~al.}(2012)\citenamefont
  {Watahiki}, \citenamefont {Shimada}, \citenamefont {Zhao}, \citenamefont
  {Chiashi}, \citenamefont {Iwamoto}, \citenamefont {Arakawa}, \citenamefont
  {Maruyama},\ and\ \citenamefont {Kato}}]{Watahiki:2012}%
  \BibitemOpen
  \bibfield  {author} {\bibinfo {author} {\bibfnamefont {R.}~\bibnamefont
  {Watahiki}}, \bibinfo {author} {\bibfnamefont {T.}~\bibnamefont {Shimada}},
  \bibinfo {author} {\bibfnamefont {P.}~\bibnamefont {Zhao}}, \bibinfo {author}
  {\bibfnamefont {S.}~\bibnamefont {Chiashi}}, \bibinfo {author} {\bibfnamefont
  {S.}~\bibnamefont {Iwamoto}}, \bibinfo {author} {\bibfnamefont
  {Y.}~\bibnamefont {Arakawa}}, \bibinfo {author} {\bibfnamefont
  {S.}~\bibnamefont {Maruyama}}, \ and\ \bibinfo {author} {\bibfnamefont
  {Y.~K.}\ \bibnamefont {Kato}},\ }\bibfield  {title} {\bibinfo {title}
  {Enhancement of carbon nanotube photoluminescence by photonic crystal
  nanocavities},\ }\href {\doibase 10.1063/1.4757876} {\bibfield  {journal}
  {\bibinfo  {journal} {Appl. Phys. Lett.}\ }\textbf {\bibinfo {volume}
  {101}},\ \bibinfo {pages} {141124} (\bibinfo {year} {2012})}\BibitemShut
  {NoStop}%
\bibitem [{\citenamefont {Imamura}\ \emph {et~al.}(2013)\citenamefont
  {Imamura}, \citenamefont {Watahiki}, \citenamefont {Miura}, \citenamefont
  {Shimada},\ and\ \citenamefont {Kato}}]{Imamura:2013}%
  \BibitemOpen
  \bibfield  {author} {\bibinfo {author} {\bibfnamefont {S.}~\bibnamefont
  {Imamura}}, \bibinfo {author} {\bibfnamefont {R.}~\bibnamefont {Watahiki}},
  \bibinfo {author} {\bibfnamefont {R.}~\bibnamefont {Miura}}, \bibinfo
  {author} {\bibfnamefont {T.}~\bibnamefont {Shimada}}, \ and\ \bibinfo
  {author} {\bibfnamefont {Y.~K.}\ \bibnamefont {Kato}},\ }\bibfield  {title}
  {\bibinfo {title} {Optical control of individual carbon nanotube light
  emitters by spectral double resonance in silicon microdisk resonators},\
  }\href {\doibase 10.1063/1.4802930} {\bibfield  {journal} {\bibinfo
  {journal} {Appl. Phys. Lett.}\ }\textbf {\bibinfo {volume} {102}},\ \bibinfo
  {pages} {161102} (\bibinfo {year} {2013})}\BibitemShut {NoStop}%
\bibitem [{\citenamefont {Miura}\ \emph {et~al.}(2014)\citenamefont {Miura},
  \citenamefont {Imamura}, \citenamefont {Ohta}, \citenamefont {Ishii},
  \citenamefont {Liu}, \citenamefont {Shimada}, \citenamefont {Iwamoto},
  \citenamefont {Arakawa},\ and\ \citenamefont {Kato}}]{Miura:2014}%
  \BibitemOpen
  \bibfield  {author} {\bibinfo {author} {\bibfnamefont {R.}~\bibnamefont
  {Miura}}, \bibinfo {author} {\bibfnamefont {S.}~\bibnamefont {Imamura}},
  \bibinfo {author} {\bibfnamefont {R.}~\bibnamefont {Ohta}}, \bibinfo {author}
  {\bibfnamefont {A.}~\bibnamefont {Ishii}}, \bibinfo {author} {\bibfnamefont
  {X.}~\bibnamefont {Liu}}, \bibinfo {author} {\bibfnamefont {T.}~\bibnamefont
  {Shimada}}, \bibinfo {author} {\bibfnamefont {S.}~\bibnamefont {Iwamoto}},
  \bibinfo {author} {\bibfnamefont {Y.}~\bibnamefont {Arakawa}}, \ and\
  \bibinfo {author} {\bibfnamefont {Y.~K.}\ \bibnamefont {Kato}},\ }\bibfield
  {title} {\bibinfo {title} {Ultralow mode-volume photonic crystal nanobeam
  cavities for high-efficiency coupling to individual carbon nanotube
  emitters},\ }\href {\doibase 10.1038/ncomms6580} {\bibfield  {journal}
  {\bibinfo  {journal} {Nat. Commun.}\ }\textbf {\bibinfo {volume} {5}},\
  \bibinfo {pages} {5580} (\bibinfo {year} {2014})}\BibitemShut {NoStop}%
\bibitem [{\citenamefont {Noury}\ \emph {et~al.}(2015)\citenamefont {Noury},
  \citenamefont {Roux}, \citenamefont {Vivien},\ and\ \citenamefont
  {Izard}}]{Noury:2015}%
  \BibitemOpen
  \bibfield  {author} {\bibinfo {author} {\bibfnamefont {A.}~\bibnamefont
  {Noury}}, \bibinfo {author} {\bibfnamefont {X.~L.}\ \bibnamefont {Roux}},
  \bibinfo {author} {\bibfnamefont {L.}~\bibnamefont {Vivien}}, \ and\ \bibinfo
  {author} {\bibfnamefont {N.}~\bibnamefont {Izard}},\ }\bibfield  {title}
  {\bibinfo {title} {Enhanced light emission from carbon nanotubes integrated
  in silicon micro-resonator},\ }\href {\doibase
  10.1088/0957-4484/26/34/345201} {\bibfield  {journal} {\bibinfo  {journal}
  {Nanotechnology}\ }\textbf {\bibinfo {volume} {26}},\ \bibinfo {pages}
  {345201} (\bibinfo {year} {2015})}\BibitemShut {NoStop}%
\bibitem [{\citenamefont {Liu}\ \emph {et~al.}(2015)\citenamefont {Liu},
  \citenamefont {Shimada}, \citenamefont {Miura}, \citenamefont {Iwamoto},
  \citenamefont {Arakawa},\ and\ \citenamefont {Kato}}]{Liu:2015}%
  \BibitemOpen
  \bibfield  {author} {\bibinfo {author} {\bibfnamefont {X.}~\bibnamefont
  {Liu}}, \bibinfo {author} {\bibfnamefont {T.}~\bibnamefont {Shimada}},
  \bibinfo {author} {\bibfnamefont {R.}~\bibnamefont {Miura}}, \bibinfo
  {author} {\bibfnamefont {S.}~\bibnamefont {Iwamoto}}, \bibinfo {author}
  {\bibfnamefont {Y.}~\bibnamefont {Arakawa}}, \ and\ \bibinfo {author}
  {\bibfnamefont {Y.~K.}\ \bibnamefont {Kato}},\ }\bibfield  {title} {\bibinfo
  {title} {Localized guided-mode and cavity-mode double resonance in photonic
  crystal nanocavities},\ }\href {\doibase 10.1103/PhysRevApplied.3.014006}
  {\bibfield  {journal} {\bibinfo  {journal} {Phys. Rev. Applied}\ }\textbf
  {\bibinfo {volume} {3}},\ \bibinfo {pages} {014006} (\bibinfo {year}
  {2015})}\BibitemShut {NoStop}%
\bibitem [{\citenamefont {Pyatkov}\ \emph {et~al.}(2016)\citenamefont
  {Pyatkov}, \citenamefont {F\"{u}tterling}, \citenamefont {Khasminskaya},
  \citenamefont {Flavel}, \citenamefont {Hennrich}, \citenamefont {Kappes},
  \citenamefont {Krupke},\ and\ \citenamefont {Pernice}}]{Pyatkov:2016}%
  \BibitemOpen
  \bibfield  {author} {\bibinfo {author} {\bibfnamefont {F.}~\bibnamefont
  {Pyatkov}}, \bibinfo {author} {\bibfnamefont {V.}~\bibnamefont
  {F\"{u}tterling}}, \bibinfo {author} {\bibfnamefont {S.}~\bibnamefont
  {Khasminskaya}}, \bibinfo {author} {\bibfnamefont {B.~S.}\ \bibnamefont
  {Flavel}}, \bibinfo {author} {\bibfnamefont {F.}~\bibnamefont {Hennrich}},
  \bibinfo {author} {\bibfnamefont {M.~M.}\ \bibnamefont {Kappes}}, \bibinfo
  {author} {\bibfnamefont {R.}~\bibnamefont {Krupke}}, \ and\ \bibinfo {author}
  {\bibfnamefont {W.~H.~P.}\ \bibnamefont {Pernice}},\ }\bibfield  {title}
  {\bibinfo {title} {Cavity-enhanced light emission from electrically driven
  carbon nanotubes},\ }\href {\doibase 10.1038/nphoton.2016.70} {\bibfield
  {journal} {\bibinfo  {journal} {Nat. Photon.}\ }\textbf {\bibinfo {volume}
  {10}},\ \bibinfo {pages} {420} (\bibinfo {year} {2016})}\BibitemShut
  {NoStop}%
\bibitem [{\citenamefont {Hoang}\ \emph {et~al.}(2017)\citenamefont {Hoang},
  \citenamefont {Dur{\'a}n-Valdeiglesias}, \citenamefont {Alonso-Ramos},
  \citenamefont {Serna}, \citenamefont {Zhang}, \citenamefont {Balestrieri},
  \citenamefont {Keita}, \citenamefont {Caselli}, \citenamefont {Biccari},
  \citenamefont {Le~Roux}, \citenamefont {Filoramo}, \citenamefont {Gurioli},
  \citenamefont {Vivien},\ and\ \citenamefont {Cassan}}]{Hoang:2017}%
  \BibitemOpen
  \bibfield  {author} {\bibinfo {author} {\bibfnamefont {T.~H.~C.}\
  \bibnamefont {Hoang}}, \bibinfo {author} {\bibfnamefont {E.}~\bibnamefont
  {Dur{\'a}n-Valdeiglesias}}, \bibinfo {author} {\bibfnamefont
  {C.}~\bibnamefont {Alonso-Ramos}}, \bibinfo {author} {\bibfnamefont
  {S.}~\bibnamefont {Serna}}, \bibinfo {author} {\bibfnamefont
  {W.}~\bibnamefont {Zhang}}, \bibinfo {author} {\bibfnamefont
  {M.}~\bibnamefont {Balestrieri}}, \bibinfo {author} {\bibfnamefont {A.-S.}\
  \bibnamefont {Keita}}, \bibinfo {author} {\bibfnamefont {N.}~\bibnamefont
  {Caselli}}, \bibinfo {author} {\bibfnamefont {F.}~\bibnamefont {Biccari}},
  \bibinfo {author} {\bibfnamefont {X.}~\bibnamefont {Le~Roux}}, \bibinfo
  {author} {\bibfnamefont {A.}~\bibnamefont {Filoramo}}, \bibinfo {author}
  {\bibfnamefont {M.}~\bibnamefont {Gurioli}}, \bibinfo {author} {\bibfnamefont
  {L.}~\bibnamefont {Vivien}}, \ and\ \bibinfo {author} {\bibfnamefont
  {E.}~\bibnamefont {Cassan}},\ }\bibfield  {title} {\bibinfo {title}
  {Narrow-linewidth carbon nanotube emission in silicon hollow-core photonic
  crystal cavity},\ }\href {\doibase 10.1364/OL.42.002228} {\bibfield
  {journal} {\bibinfo  {journal} {Opt. Lett.}\ }\textbf {\bibinfo {volume}
  {42}},\ \bibinfo {pages} {2228} (\bibinfo {year} {2017})}\BibitemShut
  {NoStop}%
\bibitem [{\citenamefont {Biccari}\ \emph {et~al.}(2017)\citenamefont
  {Biccari}, \citenamefont {Sarti}, \citenamefont {Caselli}, \citenamefont
  {Vinattieri}, \citenamefont {Dur{\'a}n-Valdeiglesias}, \citenamefont {Zhang},
  \citenamefont {Alonso-Ramos}, \citenamefont {Hoang}, \citenamefont {Serna},
  \citenamefont {Le~Roux}, \citenamefont {Cassan}, \citenamefont {Vivien},\
  and\ \citenamefont {Gurioli}}]{Biccari:2017}%
  \BibitemOpen
  \bibfield  {author} {\bibinfo {author} {\bibfnamefont {F.}~\bibnamefont
  {Biccari}}, \bibinfo {author} {\bibfnamefont {F.}~\bibnamefont {Sarti}},
  \bibinfo {author} {\bibfnamefont {N.}~\bibnamefont {Caselli}}, \bibinfo
  {author} {\bibfnamefont {A.}~\bibnamefont {Vinattieri}}, \bibinfo {author}
  {\bibfnamefont {E.}~\bibnamefont {Dur{\'a}n-Valdeiglesias}}, \bibinfo
  {author} {\bibfnamefont {W.}~\bibnamefont {Zhang}}, \bibinfo {author}
  {\bibfnamefont {C.}~\bibnamefont {Alonso-Ramos}}, \bibinfo {author}
  {\bibfnamefont {T.~H.~C.}\ \bibnamefont {Hoang}}, \bibinfo {author}
  {\bibfnamefont {S.}~\bibnamefont {Serna}}, \bibinfo {author} {\bibfnamefont
  {X.}~\bibnamefont {Le~Roux}}, \bibinfo {author} {\bibfnamefont
  {E.}~\bibnamefont {Cassan}}, \bibinfo {author} {\bibfnamefont
  {L.}~\bibnamefont {Vivien}}, \ and\ \bibinfo {author} {\bibfnamefont
  {M.}~\bibnamefont {Gurioli}},\ }\bibfield  {title} {\bibinfo {title} {Single
  walled carbon nanotubes emission coupled with a silicon slot-ring
  resonator},\ }\href {\doibase 10.1016/j.jlumin.2016.11.040} {\bibfield
  {journal} {\bibinfo  {journal} {J. Lumin.}\ }\textbf {\bibinfo {volume}
  {191}},\ \bibinfo {pages} {126} (\bibinfo {year} {2017})}\BibitemShut
  {NoStop}%
\bibitem [{\citenamefont {Machiya}\ \emph {et~al.}(2018)\citenamefont
  {Machiya}, \citenamefont {Uda}, \citenamefont {Ishii},\ and\ \citenamefont
  {Kato}}]{Machiya:2018}%
  \BibitemOpen
  \bibfield  {author} {\bibinfo {author} {\bibfnamefont {H.}~\bibnamefont
  {Machiya}}, \bibinfo {author} {\bibfnamefont {T.}~\bibnamefont {Uda}},
  \bibinfo {author} {\bibfnamefont {A.}~\bibnamefont {Ishii}}, \ and\ \bibinfo
  {author} {\bibfnamefont {Y.~K.}\ \bibnamefont {Kato}},\ }\bibfield  {title}
  {\bibinfo {title} {Spectral tuning of optical coupling between air-mode
  nanobeam cavities and individual carbon nanotubes},\ }\href {\doibase
  10.1063/1.5008299} {\bibfield  {journal} {\bibinfo  {journal} {Appl. Phys.
  Lett.}\ }\textbf {\bibinfo {volume} {112}},\ \bibinfo {pages} {021101}
  (\bibinfo {year} {2018})}\BibitemShut {NoStop}%
\bibitem [{\citenamefont {Ishii}\ \emph {et~al.}(2018)\citenamefont {Ishii},
  \citenamefont {He}, \citenamefont {Hartmann}, \citenamefont {Machiya},
  \citenamefont {Htoon}, \citenamefont {Doorn},\ and\ \citenamefont
  {Kato}}]{Ishii:2018}%
  \BibitemOpen
  \bibfield  {author} {\bibinfo {author} {\bibfnamefont {A.}~\bibnamefont
  {Ishii}}, \bibinfo {author} {\bibfnamefont {X.}~\bibnamefont {He}}, \bibinfo
  {author} {\bibfnamefont {N.~F.}\ \bibnamefont {Hartmann}}, \bibinfo {author}
  {\bibfnamefont {H.}~\bibnamefont {Machiya}}, \bibinfo {author} {\bibfnamefont
  {H.}~\bibnamefont {Htoon}}, \bibinfo {author} {\bibfnamefont {S.~K.}\
  \bibnamefont {Doorn}}, \ and\ \bibinfo {author} {\bibfnamefont {Y.~K.}\
  \bibnamefont {Kato}},\ }\bibfield  {title} {\bibinfo {title} {Enhanced
  single-photon emission from carbon-nanotube dopant states coupled to silicon
  microcavities},\ }\href {\doibase 10.1021/acs.nanolett.8b01170} {\bibfield
  {journal} {\bibinfo  {journal} {Nano Lett.}\ }\textbf {\bibinfo {volume}
  {18}},\ \bibinfo {pages} {3873} (\bibinfo {year} {2018})}\BibitemShut
  {NoStop}%
\bibitem [{\citenamefont {Dur{\'a}n-Valdeiglesias}\ \emph
  {et~al.}(2018)\citenamefont {Dur{\'a}n-Valdeiglesias}, \citenamefont {Zhang},
  \citenamefont {Alonso-Ramos}, \citenamefont {Serna}, \citenamefont {Le~Roux},
  \citenamefont {Maris-Morini}, \citenamefont {Caselli}, \citenamefont
  {Biccari}, \citenamefont {Gurioli}, \citenamefont {Filoramo}, \citenamefont
  {Cassan},\ and\ \citenamefont {Vivien}}]{DuranValdeiglesias:2018}%
  \BibitemOpen
  \bibfield  {author} {\bibinfo {author} {\bibfnamefont {E.}~\bibnamefont
  {Dur{\'a}n-Valdeiglesias}}, \bibinfo {author} {\bibfnamefont
  {W.}~\bibnamefont {Zhang}}, \bibinfo {author} {\bibfnamefont
  {C.}~\bibnamefont {Alonso-Ramos}}, \bibinfo {author} {\bibfnamefont
  {S.}~\bibnamefont {Serna}}, \bibinfo {author} {\bibfnamefont
  {X.}~\bibnamefont {Le~Roux}}, \bibinfo {author} {\bibfnamefont
  {D.}~\bibnamefont {Maris-Morini}}, \bibinfo {author} {\bibfnamefont
  {N.}~\bibnamefont {Caselli}}, \bibinfo {author} {\bibfnamefont
  {F.}~\bibnamefont {Biccari}}, \bibinfo {author} {\bibfnamefont
  {M.}~\bibnamefont {Gurioli}}, \bibinfo {author} {\bibfnamefont
  {A.}~\bibnamefont {Filoramo}}, \bibinfo {author} {\bibfnamefont
  {E.}~\bibnamefont {Cassan}}, \ and\ \bibinfo {author} {\bibfnamefont
  {L.}~\bibnamefont {Vivien}},\ }\bibfield  {title} {\bibinfo {title}
  {Tailoring carbon nanotubes optical properties through chirality-wise silicon
  ring resonators},\ }\href {\doibase 10.1038/s41598-018-29300-1} {\bibfield
  {journal} {\bibinfo  {journal} {Sci. Rep.}\ }\textbf {\bibinfo {volume}
  {8}},\ \bibinfo {eid} {11252} (\bibinfo {year} {2018})}\BibitemShut {NoStop}%
\bibitem [{\citenamefont {Zhang}\ \emph {et~al.}(2020)\citenamefont {Zhang},
  \citenamefont {Dur{\'a}n-Valdeiglesias}, \citenamefont {Alonso-Ramos},
  \citenamefont {Serna}, \citenamefont {Le~Roux}, \citenamefont {Vivien},\ and\
  \citenamefont {Cassan}}]{Zhang:2020}%
  \BibitemOpen
  \bibfield  {author} {\bibinfo {author} {\bibfnamefont {W.}~\bibnamefont
  {Zhang}}, \bibinfo {author} {\bibfnamefont {E.}~\bibnamefont
  {Dur{\'a}n-Valdeiglesias}}, \bibinfo {author} {\bibfnamefont {C.}~\bibnamefont
  {Alonso-Ramos}}, \bibinfo {author} {\bibfnamefont {S.}~\bibnamefont {Serna}},
  \bibinfo {author} {\bibfnamefont {X.}~\bibnamefont {Le~Roux}}, \bibinfo
  {author} {\bibfnamefont {L.}~\bibnamefont {Vivien}}, \ and\ \bibinfo {author}
  {\bibfnamefont {E.}~\bibnamefont {Cassan}},\ }\bibfield  {title} {\bibinfo
  {title} {Integration of semiconducting carbon nanotubes within a silicon
  photonic molecule},\ }\href {\doibase 10.1109/JPHOT.2020.2964647} {\bibfield
  {journal} {\bibinfo  {journal} {IEEE Photonics J.}\ }\textbf {\bibinfo
  {volume} {12}},\ \bibinfo {eid} {2964647} (\bibinfo {year}
  {2020})}\BibitemShut {NoStop}%
\bibitem [{\citenamefont {Higuchi}\ \emph {et~al.}(2020)\citenamefont
  {Higuchi}, \citenamefont {Niiyama}, \citenamefont {Nakagawa},\ and\
  \citenamefont {Maki}}]{Higuchi:2020}%
  \BibitemOpen
  \bibfield  {author} {\bibinfo {author} {\bibfnamefont {N.}~\bibnamefont
  {Higuchi}}, \bibinfo {author} {\bibfnamefont {H.}~\bibnamefont {Niiyama}},
  \bibinfo {author} {\bibfnamefont {K.}~\bibnamefont {Nakagawa}}, \ and\
  \bibinfo {author} {\bibfnamefont {H.}~\bibnamefont {Maki}},\ }\bibfield
  {title} {\bibinfo {title} {Efficient and narrow-linewidth photoluminescence
  devices based on single-walled carbon nanotubes and silicon photonics},\
  }\href {\doibase 10.1021/acsanm.0c01296} {\bibfield  {journal} {\bibinfo
  {journal} {ACS Appl. Nano Mater.}\ }\textbf {\bibinfo {volume} {3}},\
  \bibinfo {pages} {7678} (\bibinfo {year} {2020})}\BibitemShut {NoStop}%
\bibitem [{\citenamefont {Quan}\ \emph {et~al.}(2010)\citenamefont {Quan},
  \citenamefont {Deotare},\ and\ \citenamefont {Loncar}}]{Quan:2010}%
  \BibitemOpen
  \bibfield  {author} {\bibinfo {author} {\bibfnamefont {Q.}~\bibnamefont
  {Quan}}, \bibinfo {author} {\bibfnamefont {P.~B.}\ \bibnamefont {Deotare}}, \
  and\ \bibinfo {author} {\bibfnamefont {M.}~\bibnamefont {Loncar}},\
  }\bibfield  {title} {\bibinfo {title} {Photonic crystal nanobeam cavity
  strongly coupled to the feeding waveguide},\ }\href {\doibase
  10.1063/1.3429125} {\bibfield  {journal} {\bibinfo  {journal} {Appl. Phys.
  Lett.}\ }\textbf {\bibinfo {volume} {96}},\ \bibinfo {pages} {203102}
  (\bibinfo {year} {2010})}\BibitemShut {NoStop}%
\bibitem [{\citenamefont {Akahane}\ \emph {et~al.}(2003)\citenamefont
  {Akahane}, \citenamefont {Asano}, \citenamefont {Song},\ and\ \citenamefont
  {Noda}}]{Akahane:2003Nature}%
  \BibitemOpen
  \bibfield  {author} {\bibinfo {author} {\bibfnamefont {Y.}~\bibnamefont
  {Akahane}}, \bibinfo {author} {\bibfnamefont {T.}~\bibnamefont {Asano}},
  \bibinfo {author} {\bibfnamefont {B.-S.}\ \bibnamefont {Song}}, \ and\
  \bibinfo {author} {\bibfnamefont {S.}~\bibnamefont {Noda}},\ }\bibfield
  {title} {\bibinfo {title} {High-${Q}$ photonic nanocavity in a
  two-dimensional photonic crystal},\ }\href
  {http://dx.doi.org/10.1038/nature02063} {\bibfield  {journal} {\bibinfo
  {journal} {Nature}\ }\textbf {\bibinfo {volume} {425}},\ \bibinfo {pages}
  {944} (\bibinfo {year} {2003})}\BibitemShut {NoStop}%
\bibitem [{\citenamefont {Soltani}\ \emph {et~al.}(2007)\citenamefont
  {Soltani}, \citenamefont {Yegnanarayanan},\ and\ \citenamefont
  {Adibi}}]{Soltani:2007}%
  \BibitemOpen
  \bibfield  {author} {\bibinfo {author} {\bibfnamefont {M.}~\bibnamefont
  {Soltani}}, \bibinfo {author} {\bibfnamefont {S.}~\bibnamefont
  {Yegnanarayanan}}, \ and\ \bibinfo {author} {\bibfnamefont {A.}~\bibnamefont
  {Adibi}},\ }\bibfield  {title} {\bibinfo {title} {Ultra-high ${Q}$ planar
  silicon microdisk resonators for chip-scale silicon photonics},\ }\href
  {\doibase 10.1364/OE.15.004694} {\bibfield  {journal} {\bibinfo  {journal}
  {Opt. Express}\ }\textbf {\bibinfo {volume} {15}},\ \bibinfo {pages} {4694}
  (\bibinfo {year} {2007})}\BibitemShut {NoStop}%
\bibitem [{\citenamefont {Bogaerts}\ \emph {et~al.}(2012)\citenamefont
  {Bogaerts}, \citenamefont {De~Heyn}, \citenamefont {Van~Vaerenbergh},
  \citenamefont {De~Vos}, \citenamefont {Kumar~Selvaraja}, \citenamefont
  {Claes}, \citenamefont {Dumon}, \citenamefont {Bienstman}, \citenamefont
  {Van~Thourhout},\ and\ \citenamefont {Baets}}]{Bogaerts:2012}%
  \BibitemOpen
  \bibfield  {author} {\bibinfo {author} {\bibfnamefont {W.}~\bibnamefont
  {Bogaerts}}, \bibinfo {author} {\bibfnamefont {P.}~\bibnamefont {De~Heyn}},
  \bibinfo {author} {\bibfnamefont {T.}~\bibnamefont {Van~Vaerenbergh}},
  \bibinfo {author} {\bibfnamefont {K.}~\bibnamefont {De~Vos}}, \bibinfo
  {author} {\bibfnamefont {S.}~\bibnamefont {Kumar~Selvaraja}}, \bibinfo
  {author} {\bibfnamefont {T.}~\bibnamefont {Claes}}, \bibinfo {author}
  {\bibfnamefont {P.}~\bibnamefont {Dumon}}, \bibinfo {author} {\bibfnamefont
  {P.}~\bibnamefont {Bienstman}}, \bibinfo {author} {\bibfnamefont
  {D.}~\bibnamefont {Van~Thourhout}}, \ and\ \bibinfo {author} {\bibfnamefont
  {R.}~\bibnamefont {Baets}},\ }\bibfield  {title} {\bibinfo {title} {Silicon
  microring resonators},\ }\href {\doibase 10.1002/lpor.201100017} {\bibfield
  {journal} {\bibinfo  {journal} {Laser Photon. Rev.}\ }\textbf {\bibinfo
  {volume} {6}},\ \bibinfo {pages} {47} (\bibinfo {year} {2012})}\BibitemShut
  {NoStop}%
\bibitem [{\citenamefont {Oskooi}\ \emph {et~al.}(2010)\citenamefont {Oskooi},
  \citenamefont {Roundy}, \citenamefont {Ibanescu}, \citenamefont {Bermel},
  \citenamefont {Joannopoulos},\ and\ \citenamefont {Johnson}}]{Oskooi:2010}%
  \BibitemOpen
  \bibfield  {author} {\bibinfo {author} {\bibfnamefont {A.~F.}\ \bibnamefont
  {Oskooi}}, \bibinfo {author} {\bibfnamefont {D.}~\bibnamefont {Roundy}},
  \bibinfo {author} {\bibfnamefont {M.}~\bibnamefont {Ibanescu}}, \bibinfo
  {author} {\bibfnamefont {P.}~\bibnamefont {Bermel}}, \bibinfo {author}
  {\bibfnamefont {J.~D.}\ \bibnamefont {Joannopoulos}}, \ and\ \bibinfo
  {author} {\bibfnamefont {S.~G.}\ \bibnamefont {Johnson}},\ }\bibfield
  {title} {\bibinfo {title} {Meep: A flexible free-software package for
  electromagnetic simulations by the FDTD method},\ }\href {\doibase
  10.1016/j.cpc.2009.11.008} {\bibfield  {journal} {\bibinfo  {journal}
  {Comput. Phys. Commun.}\ }\textbf {\bibinfo {volume} {181}},\
  \bibinfo {pages} {687} (\bibinfo {year} {2010})}\BibitemShut {NoStop}%
\bibitem [{\citenamefont {Ishii}\ \emph {et~al.}(2015)\citenamefont {Ishii},
  \citenamefont {Yoshida},\ and\ \citenamefont {Kato}}]{Ishii:2015}%
  \BibitemOpen
  \bibfield  {author} {\bibinfo {author} {\bibfnamefont {A.}~\bibnamefont
  {Ishii}}, \bibinfo {author} {\bibfnamefont {M.}~\bibnamefont {Yoshida}}, \
  and\ \bibinfo {author} {\bibfnamefont {Y.~K.}\ \bibnamefont {Kato}},\
  }\bibfield  {title} {\bibinfo {title} {Exciton diffusion, end quenching, and
  exciton-exciton annihilation in individual air-suspended carbon nanotubes},\
  }\href {\doibase 10.1103/PhysRevB.91.125427} {\bibfield  {journal} {\bibinfo
  {journal} {Phys. Rev. B}\ }\textbf {\bibinfo {volume} {91}},\ \bibinfo
  {pages} {125427} (\bibinfo {year} {2015})}\BibitemShut {NoStop}%
\bibitem [{\citenamefont {Otsuka}\ \emph {et~al.}(2020)\citenamefont {Otsuka},
  \citenamefont {Fang}, \citenamefont {Yamashita}, \citenamefont {Taniguchi},
  \citenamefont {Watanabe},\ and\ \citenamefont {Kato}}]{Otsuka:2020}%
  \BibitemOpen
  \bibfield  {author} {\bibinfo {author} {\bibfnamefont {K.}~\bibnamefont
  {Otsuka}}, \bibinfo {author} {\bibfnamefont {N.}~\bibnamefont {Fang}},
  \bibinfo {author} {\bibfnamefont {D.}~\bibnamefont {Yamashita}}, \bibinfo
  {author} {\bibfnamefont {T.}~\bibnamefont {Taniguchi}}, \bibinfo {author}
  {\bibfnamefont {K.}~\bibnamefont {Watanabe}}, \ and\ \bibinfo {author}
  {\bibfnamefont {Y.~K.}\ \bibnamefont {Kato}},\ }\bibfield  {title} {\bibinfo
  {title} {Deterministic transfer of optical-quality carbon nanotubes for
  atomically defined technology},\ }\href@noop {} {\bibfield  {journal}
  {\bibinfo  {journal} {arXiv:2012.01741}\ } (\bibinfo {year}
  {2020})}\BibitemShut {NoStop}%
\bibitem [{\citenamefont {Quan}\ and\ \citenamefont
  {Loncar}(2011)}]{Quan:2011}%
  \BibitemOpen
  \bibfield  {author} {\bibinfo {author} {\bibfnamefont {Q.}~\bibnamefont
  {Quan}}\ and\ \bibinfo {author} {\bibfnamefont {M.}~\bibnamefont {Loncar}},\
  }\bibfield  {title} {\bibinfo {title} {Deterministic design of wavelength
  scale, ultra-high {Q} photonic crystal nanobeam cavities},\ }\href {\doibase
  10.1364/OE.19.018529} {\bibfield  {journal} {\bibinfo  {journal} {Opt.
  Express}\ }\textbf {\bibinfo {volume} {19}},\ \bibinfo {pages} {18529}
  (\bibinfo {year} {2011})}\BibitemShut {NoStop}%
\bibitem [{\citenamefont {Shani}\ \emph {et~al.}(1989)\citenamefont {Shani},
  \citenamefont {Henry}, \citenamefont {Kistler}, \citenamefont {Orlowsky},\
  and\ \citenamefont {Ackerman}}]{Shani:1989}%
  \BibitemOpen
  \bibfield  {author} {\bibinfo {author} {\bibfnamefont {Y.}~\bibnamefont
  {Shani}}, \bibinfo {author} {\bibfnamefont {C.~H.}\ \bibnamefont {Henry}},
  \bibinfo {author} {\bibfnamefont {R.~C.}\ \bibnamefont {Kistler}}, \bibinfo
  {author} {\bibfnamefont {K.~J.}\ \bibnamefont {Orlowsky}}, \ and\ \bibinfo
  {author} {\bibfnamefont {D.~A.}\ \bibnamefont {Ackerman}},\ }\bibfield
  {title} {\bibinfo {title} {Efficient coupling of a semiconductor laser to an
  optical fiber by means of a tapered waveguide on silicon},\ }\href {\doibase
  10.1063/1.102290} {\bibfield  {journal} {\bibinfo  {journal} {Appl. Phys.
  Lett.}\ }\textbf {\bibinfo {volume} {55}},\ \bibinfo {pages} {2389}
  (\bibinfo {year} {1989})}\BibitemShut {NoStop}%
\bibitem [{\citenamefont {Tiecke}\ \emph {et~al.}(2015)\citenamefont {Tiecke},
  \citenamefont {Nayak}, \citenamefont {Thompson}, \citenamefont {Peyronel},
  \citenamefont {de~Leon}, \citenamefont {Vuleti{\'c}},\ and\ \citenamefont
  {Lukin}}]{Tiecke:2015}%
  \BibitemOpen
  \bibfield  {author} {\bibinfo {author} {\bibfnamefont {T.~G.}\ \bibnamefont
  {Tiecke}}, \bibinfo {author} {\bibfnamefont {K.~P.}\ \bibnamefont {Nayak}},
  \bibinfo {author} {\bibfnamefont {J.~D.}\ \bibnamefont {Thompson}}, \bibinfo
  {author} {\bibfnamefont {T.}~\bibnamefont {Peyronel}}, \bibinfo {author}
  {\bibfnamefont {N.~P.}\ \bibnamefont {de~Leon}}, \bibinfo {author}
  {\bibfnamefont {V.}~\bibnamefont {Vuleti{\'c}}}, \ and\ \bibinfo {author}
  {\bibfnamefont {M.~D.}\ \bibnamefont {Lukin}},\ }\bibfield  {title} {\bibinfo
  {title} {Efficient fiber-optical interface for nanophotonic devices},\ }\href
  {\doibase 10.1364/OPTICA.2.000070} {\bibfield  {journal} {\bibinfo  {journal}
  {Optica}\ }\textbf {\bibinfo {volume} {2}},\ \bibinfo {pages} {70} (\bibinfo
  {year} {2015})}\BibitemShut {NoStop}%
\bibitem [{\citenamefont {Jimenez~Gordillo}\ \emph {et~al.}(2019)\citenamefont
  {Jimenez~Gordillo}, \citenamefont {Chaitanya}, \citenamefont {Chang},
  \citenamefont {Dave}, \citenamefont {Mohanty},\ and\ \citenamefont
  {Lipson}}]{JimenezGordillo:2019}%
  \BibitemOpen
  \bibfield  {author} {\bibinfo {author} {\bibfnamefont {O.~A.}\ \bibnamefont
  {Jimenez~Gordillo}}, \bibinfo {author} {\bibfnamefont {S.}~\bibnamefont
  {Chaitanya}}, \bibinfo {author} {\bibfnamefont {Y.-C.}\ \bibnamefont
  {Chang}}, \bibinfo {author} {\bibfnamefont {U.~D.}\ \bibnamefont {Dave}},
  \bibinfo {author} {\bibfnamefont {A.}~\bibnamefont {Mohanty}}, \ and\
  \bibinfo {author} {\bibfnamefont {M.}~\bibnamefont {Lipson}},\ }\bibfield
  {title} {\bibinfo {title} {Plug-and-play fiber to waveguide connector},\
  }\href {\doibase 10.1364/OE.27.020305} {\bibfield  {journal} {\bibinfo
  {journal} {Opt. Express}\ }\textbf {\bibinfo {volume} {27}},\ \bibinfo
  {pages} {20305} (\bibinfo {year} {2019})}\BibitemShut {NoStop}%
\end{thebibliography}
\end{document}